\documentclass[runningheads]{llncs}
\usepackage{amsmath,amssymb,amsfonts}
\usepackage{graphicx}
\usepackage{color}
\usepackage{hyperref}
\usepackage{caption}
\usepackage{enumerate}
\usepackage{float}
\usepackage{booktabs}
\usepackage[titlenumbered,ruled]{algorithm2e}
\usepackage{tabularx,booktabs,ragged2e}

\SetKwProg{Fn}{function}{ \{}{\}}

\usepackage{listings, xcolor}

\definecolor{verylightgray}{rgb}{.97,.97,.97}

\lstdefinelanguage{Solidity}{
	keywords=[1]{anonymous, assembly, assert, balance, break, call, callcode, case, catch, class, constant, continue, constructor, contract, debugger, default, delegatecall, delete, do, else, emit, event, experimental, export, external, false, finally, for, function, gas, if, implements, import, in, indexed, instanceof, interface, internal, is, length, library, log0, log1, log2, log3, log4, memory, modifier, new, payable, pragma, private, protected, public, pure, push, require, return, returns, revert, selfdestruct, send, solidity, storage, struct, suicide, super, switch, then, this, throw, true, try, typeof, using, msg.value, view, while, with, addmod, ecrecover, keccak256, mulmod, ripemd160, sha256, sha3}, 
	keywordstyle=[1]\color{blue}\bfseries,
	keywords=[2]{address, bool, byte, bytes, bytes1, bytes2, bytes3, bytes4, bytes5, bytes6, bytes7, bytes8, bytes9, bytes10, bytes11, bytes12, bytes13, bytes14, bytes15, bytes16, bytes17, bytes18, bytes19, bytes20, bytes21, bytes22, bytes23, bytes24, bytes25, bytes26, bytes27, bytes28, bytes29, bytes30, bytes31, bytes32, enum, int, int8, int16, int24, int32, int40, int48, int56, int64, int72, int80, int88, int96, int104, int112, int120, int128, int136, int144, int152, int160, int168, int176, int184, int192, int200, int208, int216, int224, int232, int240, int248, int256, mapping, string, uint, uint8, uint16, uint24, uint32, uint40, uint48, uint56, uint64, uint72, uint80, uint88, uint96, uint104, uint112, uint120, uint128, uint136, uint144, uint152, uint160, uint168, uint176, uint184, uint192, uint200, uint208, uint216, uint224, uint232, uint240, uint248, uint256, var, void, ether, finney, szabo, wei, days, hours, minutes, seconds, weeks, years},	
	keywordstyle=[2]\color{teal}\bfseries,
	keywords=[3]{block, blockhash, coinbase, difficulty, gaslimit, number, timestamp, msg, data, gas, sender, sig, msg.value, now, tx, gasprice, origin},	
	keywordstyle=[3]\color{violet}\bfseries,
	identifierstyle=\color{black},
	sensitive=false,
	comment=[l]{//},
	morecomment=[s]{/*}{*/},
	commentstyle=\color{gray}\ttfamily,
	stringstyle=\color{red}\ttfamily,
	morestring=[b]',
	morestring=[b]"
}

\begin{document}
\title{Specification is Law: Safe Creation and Upgrade of Ethereum Smart Contracts}
\titlerunning{Specification is Law: Safe Deployment of Ethereum Smart Contracts}
%
\author{Pedro Antonino \inst{1} \and Juliandson Ferreira \inst{2} \and Augusto Sampaio \inst{2} \and A. W. Roscoe \inst{1,3,4}}
\authorrunning{P. Antonino et al.}
%
\institute{The Blockhouse Technology Limited,
Oxford, UK\\ \email{pedro@tbtl.com} \and 
Centro de Informática, Universidade Federal de Pernambuco, Recife, Brazil \\ \email{jef@cin.ufpe.br, acas@cin.ufpe.br}
\and
Department of Computer Science, Oxford University, Oxford, UK \and
University College Oxford Blockchain Research Centre, Oxford, UK \\ \email{awroscoe@gmail.com}}
\maketitle              
\begin{abstract}
Smart contracts are the building blocks of the ``code is law” paradigm: the smart contract’s code indisputably describes how
its assets are to be managed - once it is created, its code is typically immutable. Faulty smart contracts present the most significant evidence against the practicality of this paradigm; they are well-documented and resulted in assets worth vast sums of money being compromised. To address this issue, the Ethereum community proposed (i) tools and processes to audit/analyse smart contracts, and (ii) design patterns implementing a mechanism to make contract code mutable. Individually, (i) and (ii) only partially address the challenges raised by the ``code is law" paradigm. In this paper, we combine elements from (i) and (ii) to create a systematic framework that moves away from ``code is law" and gives rise to a new ``specification is law" paradigm. It allows contracts to be created and upgraded but only if they meet a corresponding formal specification. The framework is centered around \emph{a trusted deployer}: an off-chain service that formally verifies and enforces this notion of conformance. 
We have prototyped this framework, and investigated its applicability to contracts implementing two widely used Ethereum standards: the ERC20 Token Standard and ERC1155 Multi Token Standard, with promising results. 

\end{abstract}

\section{Introduction}
A \emph{smart contract} is a stateful reactive program that is stored in and processed by a trusted platform, typically a blockchain, which securely executes such a program and safely stores its persistent state. 
Smart contracts were created to provide an unambiguous, automated, and secure way to manage digital assets. They are the building blocks of the ``code is law" paradigm, indisputably describing how their assets are to be managed. To implement this paradigm, many smart contract platforms - including Ethereum, the platform we focus on - disallow the code of a contract to be changed once deployed, effectively enforcing a notion of \emph{code/implementation immutability}.

Implementation immutability, however, has two main drawbacks. Firstly, contracts cannot be patched if the implementation is found to be incorrect after being deployed. There are many examples of real-world contract instances with flaws that have been exploited with astonishing sums of cryptocurrencies being taken over~\cite{dao-coindesk,dao-motherboard,Atzei17}. The ever-increasing valuation of these assets presents a significant long-standing incentive to perpetrators of such attacks. Secondly, contracts cannot be optimised. The execution of a contract function has an explicit cost to be paid by the caller that is calculated based on the contract's implementation. Platform participants would, then, benefit from contracts being updated to a functionally-equivalent but more cost-effective implementation, which is disallowed by this sort of code immutability.

To overcome this limitation, the Ethereum community has adopted the \emph{proxy pattern}~\cite{proxy} as a mechanism by which one can mimic contract upgrades. The simple application of this pattern, however, presents a number of potential issues. Firstly, the use of this mechanism allows for the patching of smart contracts but it does not address the fundamental underlying problem of correctness. Once an issue is detected, it can be patched but (i) it may be too late, and (ii) what if the patch is faulty too? Secondly, it typically gives an, arguably, unreasonable amount of power to the maintainers of this contract. These special contract users can change the contract's code with little oversight. The main flaw of such an approach is, arguably, the fact that no guarantees are enforced by this updating process; the contract implementations can change rather arbitrarily as long as the right participants have approved the change. In such a context, the ``code is law" paradigm is in fact nonexistent.

To address these issues, we propose a \emph{systematic deployment framework} that requires contracts to be formally verified before they are created and upgraded; we target the Ethereum platform and smart contracts written in Solidity. We propose a \emph{verification framework} based on the \emph{design-by-contract methodology}~\cite{Meyer92}. The specification format that we propose is similar to what the community has used, albeit in an informal way, to specify the behaviour of common Ethereum contracts~\cite{erc20}. 
Our framework also relies on our own version of the proxy pattern to carry out updates but in a sophisticated and safe way. We rely on a \emph{trusted deployer}, which is an off-chain service, 
to vet contract creations and updates. These operations are only allowed if the given implementation meets the expected specification - the contract specification is set at the time of contract creation and remains unchanged during its lifetime. As an off-chain service, our framework can be readily and efficiently integrated into existing blockchain platforms, but the same is not true of an on-chain implementation; we elaborate on this trade-off in Section~\ref{sec:safe-deployment}. Participants can also check whether a contract has been deployed via our framework so that they can be certain the contract they want to execute has the expected behaviour. 

Our framework promotes a paradigm shift where the specification is immutable instead of the implementation/code. Thus, it moves away from ``code is law" and proposes the \emph{``specification is law" paradigm} - enforced by formal verification. This new paradigm addresses all the concerns that we have highlighted: arbitrary code updates are forbidden as only conforming implementations are allowed, and buggy contracts are prevented from being deployed as they are vetted by a formal verifier. Thus, contracts can be optimised and changed to meet evolving business needs and yet contract stakeholders can rely on the guarantee that the implementations always conform to their corresponding specifications. As specifications are more stable and a necessary element for assessing the correctness of a contract, we believe that a framework that focuses on this key artifact and makes it immutable improves on the current ``code is law" paradigm. We further discuss the benefits of this paradigm shift in Section~\ref{sec:conclusion}.

We have created a prototype of our framework, and conducted a case study that investigates its applicability to real-world smart contracts implementing the widely used ERC20 and ERC1155 Ethereum token standards~\cite{erc20,erc1155}. We analysed specifically how the sort of formal verification that we use fares in handling practical contracts and obtained promising results.

In this paper, we assume the deployer is a trusted third party and focus on the functional aspect of our framework. We are currently working on an implementation of the trusted deployer that relies on a Trusted Execution Environment (TEE)~\cite{Maene18}, specifically the AMD SEV implementation~\cite{sev20}. 

\noindent
\textbf{Outline.} Section~\ref{sec:background} introduces the relevant background material. Section~\ref{sec:safe-deployment} introduces our framework, and Section~\ref{sec:evaluation} the evaluation that we conducted. Section~\ref{sec:related-work} discusses related work, whereas Section~\ref{sec:conclusion} presents our concluding remarks.

\section{Background}
\label{sec:background}

\subsection{Ethereum and Solidity}

The Ethereum blockchain is arguably the most popular smart contract platform. A \emph{participant} in Ethereum controls \emph{addresses} in the blockchain, each of which has a balance of Ether - Ethereum's cryptocurrency - associated with it. These addresses are akin to account numbers in traditional banking. A participant that controls an address also controls the balance associated to it. Thus, they can send a \emph{transaction} to Ethereum requesting the transfer of some amount of the balance associated to one of its addresses. Aside from these addresses that are managed by external entities, Ethereum also allows addresses to be managed by a \emph{program} (a smart contract). In addition to a balance, these \emph{smart contract} addresses have some code and data associated to them. While the former defines the functions offered by the contract, the latter captures its persistent state. For a detailed presentation of Ethereum, see, for instance, \cite{EthereumWhite,EthereumYellow}. 

Solidity is arguably the most used language for writing smart contracts. A contract in Solidity is a concept very similar to that of a \emph{class} in object-oriented languages, and a contract instance a sort of long-lived persistent object. We introduce the main elements of Solidity using the \lstinline|ToyWallet| contract in Figure~\ref{fig:toy-wallet-contract}. It implements a very basic ``wallet" contract that participants and other contracts can rely upon to store their Ether. The \emph{member variables} of a contract define the persistent state of the contract. This example contract has a single member variable \lstinline|accs|, a mapping from addresses to 256-bit unsigned integers, which keeps track of the balance of Ether each ``client" of the contract has in the \lstinline|ToyWallet|; the integer \lstinline|accs[addr]| gives the current balance for address \lstinline|addr|, and an address is represented by a 160-bit number. 

Public functions describe the operations that participants and other contracts can execute on the contract. The contract in Figure~\ref{fig:toy-wallet-contract} has \emph{public functions} \lstinline|deposit| and \lstinline|withdraw| that can be used to transfer Ether into and out of the \lstinline|ToyWallet| contract, respectively. In Solidity, functions have the implicit argument \lstinline|msg.sender| designating the caller's address, and \lstinline|payable| functions have the \lstinline|msg.value| which depict how much \emph{Wei} - the most basic (sub)unit of Ether - is being transferred, from caller to callee, with that function invocation; such a transfer is carried out implicitly by Ethereum. For instance,  when \lstinline|deposit| is called on an instance of \lstinline|ToyWallet|, the caller can decide on some amount \lstinline|amt| of Wei to be sent with the invocation. By the time the \lstinline|deposit| body is about to execute, Ethereum will already have carried out the transfer from the balance associated to the caller's address to that of the \lstinline|ToyWallet| instance - and \lstinline|amt| can be accessed via \lstinline|msg.value|. Note that, as mentioned, this balance is part of the blockchain's state rather than an explicit variable declared by the contract's code. One can programmatically access this implicit balance variable for address \lstinline|addr| with the command \lstinline|addr.balance|. Solidity's construct \lstinline|require(condition)| aborts and reverts the execution of the function in question if \lstinline|condition| does not hold - even in the case of implicit Ether transfers. The call \lstinline|addr.send(amount)| sends \lstinline|amount| Wei from the currently executing instance to address \lstinline|addr|; it returns \lstinline|true| if the transfer was successful, and \lstinline|false| otherwise. For instance, the first \lstinline|require| statement in the function \lstinline|withdraw| requires the caller to have the funds they want to withdraw, whereas the second requires the \lstinline|msg.sender.send(value)| statement to succeed, i.e. the \lstinline|value| must have been correctly withdrawn from \lstinline|ToyWallet| to \lstinline|msg.sender|. The final statement in this function updates the account balance of the caller (i.e. \lstinline|msg.sender|) in \lstinline|ToyWallet| to reflect the withdrawal.

We use the transaction $\textit{create-contract}$ as a means to create an instance of a Solidity smart contract in Ethereum. In reality, Ethereum only accepts contracts in the \emph{EVM bytecode} low-level language - Solidity contracts need to be compiled into that. The processing of a transaction $\textit{create-contract}(c, args)$ creates an instance of contract $c$ and executes its constructor with arguments $args$%
. Solidity contracts without a constructor (as our example in Figure~\ref{fig:toy-wallet-contract}) are given an implicit one. A \textit{create-contract} call returns the address at which the contract instance was created. We omit the \textit{args} when they are not relevant for a call. We use $\sigma$ to denote the state of the blockchain where $\sigma[ad].balance$ gives the  balance for address $ad$, and $\sigma[ad].storage.mem$ the value for member variable $mem$ of the contract instance deployed at $ad$ for this state. For instance, let $c_{\texttt{tw}}$ be the code in Figure~\ref{fig:toy-wallet-contract}, and $\texttt{addr}_{\texttt{tw}}$ the address returned by the processing of $\textit{create-contract}(c_{\texttt{tw}})$. For the blockchain state $\sigma'$ immediately after this processing, we have that: for any address $addr$, $\sigma'[\texttt{addr}_{\texttt{tw}}].storage.accs[addr] = 0$ and its balance is zero, i.e., $\sigma'[addr_{\texttt{tw}}].balance = 0$. We introduce and use this intuitive notation to present and discuss state changes as it can concisely and clearly capture them. There are many works that formalise such concepts \cite{Hildenbrandt18,Wang18,Antonino21}.

\begin{figure}[t]
\begin{lstlisting}[basicstyle=\scriptsize\ttfamily]
contract ToyWallet {
	mapping (address => uint) accs;
	
	function deposit () payable public {
		accs[msg.sender] = accs[msg.sender] + msg.value;
	}
	
	function withdraw (uint value) public {
		require(accs[msg.sender] >= value);
		bool ok = msg.sender.send(value);
		require(ok);
		accs[msg.sender] = accs[msg.sender] - value;
	}
}	
\end{lstlisting}
\caption{ToyWallet contract example.}
\label{fig:toy-wallet-contract}
\end{figure}

A transaction $\textit{call-contract}$ can be used to invoke contract functions; processing $\textit{call-contract}(addr, \textit{func\_sig}, args)$ executes the function with signature \textit{func\_sig} at address \textit{addr} with input arguments \textit{args}. When a contract is created, the code associated with its non-constructor public functions is made available to be called by such transactions. The constructor function is only run (and available) at creation time. For instance, let $\texttt{addr}_{\texttt{tw}}$ be a fresh \lstinline|ToyWallet| instance and \lstinline|ToyWallet.deposit| give the signature of the corresponding function in Figure~\ref{fig:toy-wallet-contract}, processing the transaction $\textit{call-contract}(\texttt{addr}_{\texttt{tw}},$ \lstinline|ToyWallet.deposit|$,\linebreak args)$ where $args = \{msg.sender = \texttt{addr}_{snd}, msg.value = 10\}$ would cause the state of this instance to be updated to $\sigma''$ where we have that $\sigma''[\texttt{addr}_{\texttt{tw}}].\allowbreak{}storage.accs[\texttt{addr}_{snd}] = 10$ and $\sigma''[\texttt{addr}_{\texttt{tw}}].balance = 10$. So, the above transaction has been issued by address $\texttt{addr}_{snd}$ which has transferred 10 Wei to $\texttt{addr}_{tw}$. 

\subsection{Formal verification with \emph{solc-verify}}

The modular verifier \emph{solc-verify}~\cite{Hajdu19,Hajdu20} was created to help developers to formally check that their Solidity smart contracts behave as expected. Input contracts are manually annotated with contract \emph{invariants} and their functions with \emph{pre-} and \emph{postconditions}. An annotated Solidity contract is then translated into a Boogie program which is verified by the Boogie verifier~\cite{Barnett05,Leino08}. Its modular nature means that \emph{solc-verify} verifies functions locally/independently, and function calls are abstracted by the corresponding function's specification, rather than their implementation being precisely analysed/executed. These specification constructs have their typical meaning. An invariant is valid if it is established by the constructor and maintained by the contract's public functions, and a function meets its specification if and only if from a state satisfying its pre-conditions, any state successfully terminating respects its postconditions. So the notion is that of partial correctness. Note that an aborted and reverted execution, such as one triggered by a failing \lstinline|require| command, does not successfully terminate. We use Figure~\ref{fig:toy-wallet-withdraw-spec} illustrates a \emph{solc-verify} specification for an alternative version of the \lstinline|ToyWallet|'s \lstinline|withdraw| function. The postconditions specify that the balance of the instance and the wallet balance associated with the caller must decrease by the withdrawn amount and no other wallet balance must be affected by the call. 

\begin{figure}[t]
\begin{lstlisting}[basicstyle=\scriptsize\ttfamily]
/**
* @notice postcondition address(this).balance == __verifier_old_uint(address(this).balance) - value
* @notice postcondition accs[msg.sender] == __verifier_old_uint(accs[msg.sender]) - value
* @notice postcondition forall (address addr) addr == msg.sender || __verifier_old_uint(accs[addr]) == accs[addr]
*/
function withdraw (uint value) public {
	require(accs[msg.sender] >= value);
	bool ok;
	(ok,) = msg.sender.call.value(value)("");
	require(ok);
	accs[msg.sender] = accs[msg.sender] - value;
}
\end{lstlisting}
\caption{\lstinline|ToyWallet| alternate buggy \lstinline|withdraw| implementation with specification.}
\label{fig:toy-wallet-withdraw-spec}
\end{figure}

This alternative implementation uses \lstinline|msg.sender.call.value(value)("")|\linebreak{} instead of \lstinline|msg.sender.send(value)|. While the latter only allows for the transfer of \lstinline|value| Wei from the instance to address \lstinline|msg.sender|, the former \emph{delegates control} to \lstinline|msg.sender| in addition to the transfer of \lstinline|value|.\footnote{In fact, the function \lstinline|send| also delegates control to \lstinline|msg.sender| but it does in such a restricted way that it cannot perform any relevant computation. So, for the purpose of this paper and to simplify our exposition, we ignore this delegation.} If \lstinline|msg.sender| is a smart contract instance that calls \lstinline|withdraw| again during this control delegation, it can withdraw all the funds in this alternative \lstinline|ToyWallet| instance - even the funds that were not deposited by it. This \emph{reentrancy} bug is detected by \emph{solc-verify} when it analyses this alternative version of the contract. A similar bug was exploited in what is known as the DAO attack/hack to take over US\$53 million worth of Ether~\cite{dao-coindesk,dao-motherboard,Atzei17}. 

\section{Safe Ethereum Smart Contracts Deployment}
\label{sec:safe-deployment}

We propose a framework for the \emph{safe creation and upgrade of smart contracts} based around a \emph{trusted deployer}. This entity is trusted to only create or update contracts that have been verified to meet their corresponding specifications. A smart contract development process built around it prevents developers from deploying contracts that have not been implemented as intended. Thus, stakeholders can be sure that contract instances deployed by this entity, even if their code is upgraded, comply with the intended specification.

Our trusted deployer targets the Ethereum platform, and we implement it as an off-chain service. Generally speaking, a trusted deployer could be implemented as a smart contract in a blockchain platform, as part of its consensus rules, or as an off-chain service. In Ethereum, implementing it as a smart contract is not practically feasible as a verification infrastructure on top of the EVM~\cite{EthereumWhite} would need to be created. Furthermore, blocks have an upper limit on the computing power they can use to process their transactions, and even relatively simple computing tasks can exceed this upper limit~\cite{Wust20}. As verification is a notoriously complex computing task, it should exceed this upper limit even for reasonably small systems. Neither can we change the consensus rules for Ethereum.

\begin{figure}[t]
	\centering
	\includegraphics[scale=0.5]{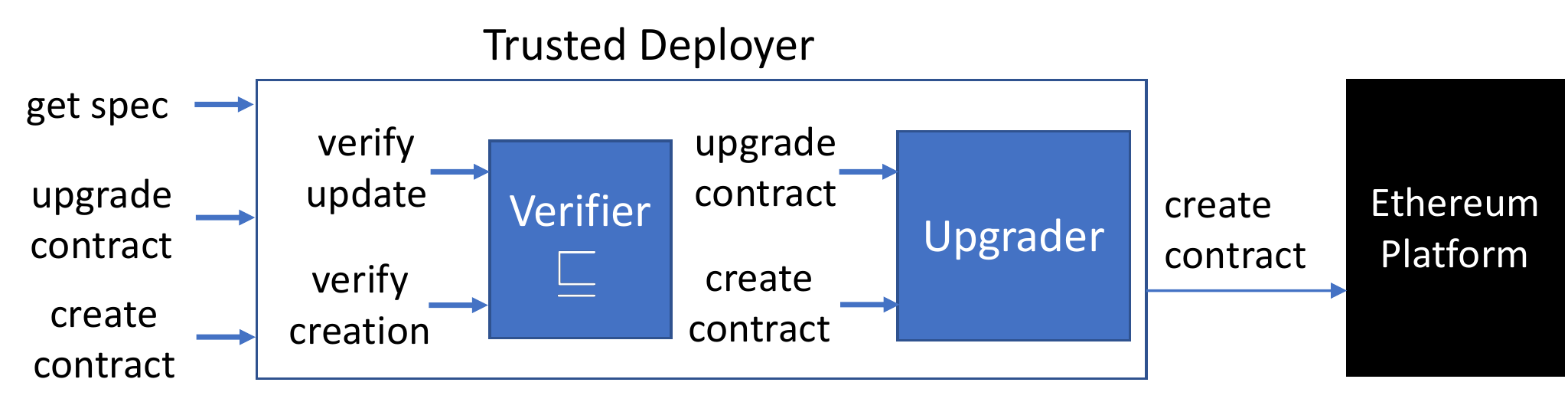}
	\caption{Trusted deployer architecture.}
	\label{fig:trusted-deployer}
\end{figure}

We present the architecture of the \emph{trusted deployer infrastructure} in Figure~\ref{fig:trusted-deployer}. The trusted deployer relies on an internal \emph{verifier} that implements the functions $\textit{verify-creation}_{\sqsubseteq}$ and $\textit{verify-upgrade}_{\sqsubseteq}$, and an \emph{upgrader} that implements functions \textit{create-contract} and \textit{upgrade-contract}; we detail what these functions do in the following. The deployer's \textit{create-contract} (\textit{upgrate-contract}) check that an implementation meets its specification by calling $\textit{verify-creation}_{\sqsubseteq}$ ($\textit{verify-upgrade}_{\sqsubseteq}$) before relaying this call to the upgrader's \textit{create-contract} (\textit{upgrade-contract}) which effectively creates (upgrades) the contract in the Ethereum platform. The \textit{get-spec} function can be used to test whether a contract instance has been deployed by the trusted deployer and which specification it satisfies.   

The \emph{verifier} is used to establish whether an implementation meets a specification. A verification framework is given by a triple $(\mathcal{S},\mathcal{C}, \sqsubseteq)$ where $\mathcal{S}$ is a language of smart contract specifications, $\mathcal{C}$ is a language of implementations, and $\sqsubseteq \in (\mathcal{S} \times \mathcal{C})$ is a satisfiability relation between smart contracts' specifications and implementations. In this paper, $\mathcal{C}$ is the set of Solidity contracts and $\mathcal{S}$ a particular form of Solidity contracts, possibly annotated with contract invariants, that include function signatures annotated with postconditions. The functions $\textit{verify-creation}_{\sqsubseteq}$ and $\textit{verify-upgrade}_{\sqsubseteq}$ both take a specification $s \in \mathcal{S}$ and a contract implementation $c \in \mathcal{C}$ and test whether $c$ meets $s$ - they work in slightly different ways as we explain later. When an implementation does not meet a specification, verifiers typically return an error report that points out which parts of the specification do not hold and maybe even witnesses/counterexamples describing system behaviours illustrating such violations; they provide valuable information to help developers correct their implementations.

The upgrader is used to create and manage \emph{upgradable smart contracts} - Ethereum does not have built-in support for contract upgrades. Function \textit{create-contract} creates an upgradable instance of contract $c \in \mathcal{C}$ - it returns the Ethereum address where the instance was created -  whereas \textit{upgrade-contract} allows for the contract's behaviour to be upgraded. The specification used for a successful contract creation will be stored and used as the intended specification for future upgrades. Only the creator of a \emph{trusted contract} can update its implementation.

Note that once a contract is created via our trusted deployer, the instance's specification is fixed, and not only its initial implementation but all upgrades are guaranteed to satisfy this specification. Therefore, participants in the ecosystem interacting with this contract instance can be certain that its behaviour is as intended by its developer during the instance's entire lifetime, even if the implementation is upgraded as the contract evolves. 

In this paper, we focus on contract upgrades that preserve the signature of public functions. Also, we assume contract specifications fix the data structures used in the contract implementation. However, we plan to relax these restrictions in future versions of the framework, allowing the data structures in the contract implementation to be a data refinement of those used in the specification; we also plan to allow the signature of the implementation to extend that of the specification, provided some notion of behaviour preservation is obeyed when the extended interface is projected into the original one.

\subsection{Verifier}

We propose \emph{design-by-contract} \cite{Meyer92} as a methodology to specify the behaviour of smart contracts. In this traditional specification paradigm, conceived for object-oriented languages, a developer can specify invariants for a class and pre-/postconditions for its methods. Invariants must be established by the constructor and guaranteed by the public methods, whereas postconditions are ensured by the code in the method's body provided that the pre-conditions are guaranteed by the caller code and the method terminates - currently, we focus on partial correctness. We propose a specification format that defines what the member variables and signatures of member functions should be. Additionally, the function signatures can be annotated with postconditions, and the specification with invariants; these annotations capture the expected behaviour of the contract. In ordinary programs, a function is called in specific \emph{call sites} fixed in the program's code. Pre-conditions can, then, be enforced and checked in these call sites. In the context of public functions of smart contracts, however, any well-formed transaction can be issued to invoke such a function. Hence, we move away from preconditions in our specification, requiring, thus, postconditions to be met whenever public functions successfully terminate.

\begin{figure}[!t]
\begin{lstlisting}[basicstyle=\scriptsize\ttfamily,  escapechar=`]
/**
* @notice invariant accs[address(this)] == 0
*/
contract ToyWallet {
	mapping (address => uint) accs; 
	
	/**
	* @notice postcondition forall (address addr) accs[addr] == 0
	*/
	constructor() public;
	
	/**
	* @notice postcondition address(this).balance == __verifier_old_uint(address(this).balance) + msg.value
	* @notice postcondition accs[msg.sender] == __verifier_old_uint(accs[msg.sender]) + msg.value
	* @notice postcondition forall (address addr) addr == msg.sender || __verifier_old_uint(accs[addr]) == accs[addr]
	*/
	function deposit () payable public;
	
	/**
	* @notice postcondition address(this).balance == __verifier_old_uint(address(this).balance) - value
	* @notice postcondition accs[msg.sender] == __verifier_old_uint(accs[msg.sender]) - value
	* @notice postcondition forall (address addr) addr == msg.sender || __verifier_old_uint(accs[addr]) == accs[addr]
	*/
	function withdraw (uint value) public;
}
\end{lstlisting}
\caption{ToyWallet specification.}
\label{fig:toy-wallet-specification}
\end{figure}

Figure~\ref{fig:toy-wallet-specification} illustrates a specification for the \lstinline|ToyWallet| contract. Invariants are declared in a comment block preceding the contract declaration, and function postconditions are declared in comment blocks preceding their signatures. Our specification language reuses constructs from Solidity and the \emph{solc-verify} specification language, which in turn borrows elements from the Boogie language~\cite{Barnett05,Leino08}. Member variables and function signature declarations are as prescribed by Solidity, whereas the conditions on invariants, and postconditions are side-effect-free Solidity expressions extended with quantifiers and the expression \lstinline|__verifier_old_|$x$\lstinline|(v)| that can only be used in a postcondition, and it denotes the value of \lstinline|v| in the function's execution pre-state.

We choose to use Solidity as opposed to EVM bytecode as it gives a cleaner semantic basis for the analysis of smart contracts~\cite{Antonino21} and it also provides a high-level error message when the specification is not met. The satisfiability relation $\sqsubseteq$ that we propose is as follows.

\begin{definition}\label{def:satisfaction} The relation $s \sqsubseteq c$ holds iff:
	\begin{itemize}
		\item \emph{Syntactic obligation}: a member variable is declared in $s$ if and only if it is declared in $c$ with the same type, and they must be declared in the same order. A public function signature is declared in $s$ if and only if it is declared and implemented in $c$.
		\item \emph{Semantic obligation}: invariants declared in $s$ must be respected by $c$, and the implementation of functions in $c$ must respect their corresponding guards and postconditions described in $s$.
	\end{itemize}
\end{definition}

The purpose of this paper is not to provide a formal semantics to Solidity or to formalise the execution model implemented by the Ethereum platform. Other works propose formalisations for Solidity and Ethereum~\cite{Hajdu19,Antonino20,Wang18}. Our focus is on using the modular verifier \emph{solc-verify} to discharge the semantic obligations imposed by our satisfaction definition.

The $\textit{verify-creation}_{\sqsubseteq}$ function works as follows. Firstly, the syntactic obligation imposed by Definition \ref{def:satisfaction} is checked by a syntactic comparison between $s$ and $c$. If it holds, we rely on \emph{solc-verify} to check whether the semantic obligation is fulfilled. We use what we call a \emph{merged contract} as the input to \emph{solc-verify} - it is obtained by annotating $c$ with the corresponding invariants and postconditions in $s$. If \emph{solc-verify} is able to discharge all the proof obligations associated to this merged contract, the semantic obligations are considered fulfilled, and $\textit{verify-creation}_{\sqsubseteq}$ succeeds.

Function $\textit{verify-upgrade}_{\sqsubseteq}$ is implemented in a very similar way but it relies on a slightly different satisfiability relation and merged contract. While $\textit{verify-creation}_{\sqsubseteq}$ checks that the obligations of the constructor are met by its implementation, $\textit{verify-upgrade}_{\sqsubseteq}$ \emph{assumes} they do, since the constructor is only executed - and, therefore, its implementation checked for satisfiability - at creation time. The upgrade process only checks conformance for the implementation of the (non-constructor) public functions. 

\subsection{Upgrader}
\label{sec:upgrader}

Ethereum does not provide a built-in mechanism for upgrading smart contracts. However, one can simulate this functionality using the \emph{proxy pattern}~\cite{proxy}, which splits the contract across two instances: the \emph{proxy instance} holds the persistent state and the upgrade logic, and rely on the code in an \textit{implementation instance} for its business logic. The proxy instance is the \emph{de-facto} instance that is the target of calls willing to execute the upgradable contract. It stores the address of the implementation instance it relies upon, and the behaviour of the proxy's public functions can be upgraded by changing this address. Our \emph{upgrader} relies on our own version of this pattern to deploy \emph{upgradable} contracts.

Given a contract $c$ that meets its specification according to Definition~\ref{def:satisfaction}, the upgrader creates the Solidity contract $proxy(c)$ as follows. It has the same member variable declarations, in the same order, as $c$ - having the same order is an implementation detail that is necessary to implement the sort of delegation we use as it enforces proxy and implementation instances to share the same memory layout. In addition to those, it has a new \emph{address} member variable called \lstinline|implementation| - it stores the address of the implementation instance. The constructor of $proxy(c)$ extends the constructor of $c$ with an initial setting up of the variable \lstinline|implementation|.\footnote{Instead of using the proxy pattern \lstinline|initialize| function to initialise the state of the proxy instance, we place the code that carries out the desired initialisation directly into the proxy's constructor. Our approach benefits from the inherent behaviour of constructors - which only execute once and at creation time - instead of having to implement this behaviour for the non-constructor function \lstinline|initialize|. Our Trusted Deployer, available at \url{https://github.com/stanis18/safeeveolution}, automatically generates the code for such a proxy.} This proxy contract also has a public function \lstinline|upgrade| that can be used to change the address of the implementation instance. The trusted deployer is identified by a trusted Ethereum address $\texttt{addr}_{td}$. This address is used to ensure calls to \lstinline|upgrade| can \emph{only} be issued by the trusted deployer. In the process of creating and upgrading contracts the trusted deployer acts as an external participant of the Ethereum platform. We assume that the contract implementations and specifications do not have member variables named \lstinline|implementation|, or functions named \lstinline|upgrade| to avoid name clashes.

The proxy instance relies on the low-level \lstinline|delegatecall| Solidity command to dynamically borrow and execute the function implementations defined in the contract instance at \lstinline|implementation|. When the contract instance at address \lstinline|proxy| executes \lstinline[mathescape]|implementation.delegatecall($sig$, args)|, it executes the code associated with the function with signature \textit{sig} stored in the instance at address \lstinline|implementation| but applied to the \lstinline|proxy| instance - modifying its state - instead of \lstinline|implementation|. For each (non-constructor) public function in $c$ with signature \textit{sig}, $proxy(c)$ has a corresponding function declaration whose implementation relies on \lstinline|implementation.delegatecall(|\textit{sig}\lstinline|, args)|. This command was proposed as a means to implement and deploy contracts that act as a sort of dynamic library. Such a contract is deployed with the sole purpose of other contracts borrowing and using their code.

The upgrader function $\textit{create-contract}(c)$ behaves as follows. Firstly, it issues transaction $\textit{create-contract}(c,args)$ to the Ethereum platform to create the initial implementation instance at address \texttt{addr}$_{\textit{impl}}$. Secondly, it issues transaction $\textit{create-contract}(proxy(c),args)$, such that \lstinline|implementation| would be set to \texttt{addr}$_{\textit{impl}}$, to create the proxy instance at address $\texttt{addr}_{px}$. Note that both of these transactions are issued by and using the trusted deployer's address $\texttt{addr}_{td}$. The upgrader function $\textit{upgrade-contract}(c)$ behaves similarly, but the second step issues transaction $\textit{call-contract}(\texttt{addr}_{px},$\lstinline|upgrade|$,args)$, triggering the execution of function \lstinline|upgrade| in the proxy instance and changing its \lstinline|implementation| address to the new implementation instance.

\section{Case Studies: ERC20 and ERC1155}
\label{sec:evaluation}

The presentation of our case studies is structured into three main sections: first we describe the context (Section~\ref{sec:context}), next we provide the results (Section~\ref{sec:results}), then we discuss the threats to validity and the limitations (Section~\ref{sec:threatslimitations}) and, finally, we analyze a scenario based on 0xMonorepo repository in order to find bugs included in the development process (Section~\ref{sec:upgradertool}).

\subsection{Context}
\label{sec:context}

The main purpose of the study is to provide evidence that the proposed framework can be used to ensure the deployment and evolution of smart contracts in a secure manner. The framework must be integrated into a development process and whenever there is any change in the code,
one must verify if the change meets the specification, so we intend to answer the following research question:
\textit{Does the proposed framework improve on the process of creation and evolution of smart contracts?}

In the first phase, a review of the literature was carried out. The objective was to explore the main features of smart contract development patterns and the most common error types. As a result, we could identify opportunities for the application of the framework in line with the objectives of the study. In the second phase, we identified, documented and validated requirements related to two Ethereum smart contract specifications: ERC20 and ERC1155. Each of these standards defines a contract interface and is accompanied by an informal description of its functions' behaviours. We chose these standards because they are widely used and in an advanced state of maturity. By structuring the existing requirements in natural language, we were able to extract the formal properties used in the verification process and create a corresponding formal contract specification. Then, we conducted a quantitative analysis, in order to verify the feasibility of the framework, to evaluate its effectiveness, which can be measured by the number of errors found or safe evolutions that have been proven correct, and efficiency, measured by the time to process the verification.  

We have implemented a tool to check the syntactic and semantic obligations imposed by our framework. It relies on the abstract syntax tree generated by the Solidity compiler~\cite{solc} to check the syntactic obligations, and to create a contract resulting from merging the solidity smart contract with the corresponding formal specification; this is then checked by the \emph{solc-verify} as a means to discharge the semantic obligations. We applied our framework to a number of real-world Solidity smart contract
samples implementing the ERC20 and the ERC1155 token standards. The contract samples we analysed were extracted from 12 github repositories that were public, and presented reasonably complex commit histories that changed the smart contract behavior. The samples also cover aspects of evolution that are related to improving the readability and maintenance of the code, but also optimisations where, for instance, redundant checks executed by a function were removed. The evaluation was carried out on a Lenovo IdeapadGaming3i with the operational system Windows 10, Intel(R) Core(TM) i7-10750 CPU @ 2.60GHz, 8GB of RAM, with Docker Engine 20.15.5 and Solidity compiler version 0.5.17. 

\begin{table}[!t]
\setlength{\tabcolsep}{3pt}
\resizebox{\textwidth}{!}{%
\begin{tabular}{|c|c|c|c||c|c|c|c|}
\hline
 \multicolumn{8}{|c|}{ERC20} \\
\hline
 Repository & Commit &  Time & Output & Repository & Commit &  Time & Output\\
\hline
    0xMonorepo  & 548fda & 2.85s  & WOP &  DsToken  & 3c436c & 3.77s  &  No errors\\
    0xMonorepo  & 6f2cb6 & 2.84s  & No errors &  DsToken  & 733e5c & 3.81s  &  No errors\\
    0xMonorepo  & c84be8 & 2.57s  & No errors &  DsToken  & 8b8263 & 3.08s  &  No errors\\
    Ambrosus  & 9fb24b & 3.15s  & No errors &  Klenergy & 3d4d62 & 5.14s  & No errors\\
    Ambrosus  & b1806b & 2.99s & No errors &  Klenergy  & 60263d & 1.70s  & VRE\\
    Ambrosus  & db3ea0 & 3.74s  & No errors &  OpenZeppelin  & 3a5da7 & 3.59s  & No errors\\
    
    DigixDao  & 0550e8 & 5.97s  & No errors & OpenZeppelin  & \underline{43ebb4} & 3.57s  & No errors\\
    DigixDao  & 1c0c4f & 8.82s  & No errors &  OpenZeppelin  & 5db741 & 3.87s  & No errors\\
    DigixDao  & \underline{5aee64} & 7.60s  & NTI &  OpenZeppelin  & 5dfe72 & 3.96s  & No errors\\
    DigixDao  & 6bddc6 & 7.74s  & No errors &  OpenZeppelin  & \underline{9b3710} & 3.45s  & No errors\\
    DigixDao  & 845F03 & 9.17s  & No errors &  Uniswap	 & 4e4546 & 3.67s  & No errors\\
    DigixDao  & aabf24 & 2.97s  & No errors &  Uniswap	 & \underline{55ae25} & 3.43s  & WOP\\
    DigixDao  & e221ff & 9.21s  & No errors &  Uniswap	 &e382d7 & 3.57s  & IOU\\
    DigixDao  & e320a2 & 8.89s  & No errors &  SkinCoin  & 25db99 & 0.99s  & NTI\\
    DsToken  & 08412f & 4.14s  & WOP &  SkinCoin  & 27c298 & 1.94s  & NTI\\
    DsToken  & 10c964 & 3.66s  & No errors &  SkinCoin  & ac33d8 & 3.23s  &  No errors\\
    
\hline
\multicolumn{8}{c}{} \\
\hline
 \multicolumn{8}{|c|}{ERC1155} \\
\hline
    0xSequence  & \underline{319740} & 4.82s  & No errors &  OpenZeppelin  & 0db76e & 5.59s  &  No errors\\
    0xSequence  & 578d46 & 5.31s  & No errors &  OpenZeppelin  & 440b65 & 6.61s  &  No errors\\
    0xSequence  & \underline{99012f} & 5.59s  & No errors &  OpenZeppelin  & 5db741 & 6.70s  &  No errors\\
    0xSequence  & acfa7c & 5.81s  & No errors &  OpenZeppelin  & 956d66 & 8.58s  &  No errors\\
    Desc-Stock  & 44464c & 4.34s  & IOU &  Ejin-Erc  & 30dba0 & 4.13s  & No errors\\
    Desc-Stock  & 4c5d80 & 5.18s  & IOU &  Ejin-Erc  & 614714 & 4.49s  & No errors\\
    Desc-Stock  & \underline{96d5b2} & 4.33s  & WOP &  Ejin-Erc  & bf4d04 & 4.01s  & No errors\\
    Desc-Stock  & ae8a13 & 4.24s  & IOU &  Ejin-Erc  & cc96af & 4.57s  & No errors\\
    Desc-Stock  & bf2c1a & 3.60s  & IOU &  Ejin-Erc  & e20fc9 & 3.77s  & No errors\\
    
 \hline
\end{tabular}}
\caption{ERC20 and ERC1155 Results}
\label{tab:erc20_erc1155_results}
\end{table}

\subsection{Results}
\label{sec:results}

Our framework was able to identify errors of the following categories: Integer Overflow and Underflow (IOU); Nonstandard Token Interface (NTI), when the contract does not meet the syntactic restriction defined by the standard; wrong operator (WOP), for instance, when the $<$ operator would be expected but $\leq$ is used instead; and Verification Error (VRE), when the verification process cannot be completed or the results were inconclusive. It also established conformance for some of the samples analysed. Table~\ref{tab:erc20_erc1155_results} shows the the complete list of all results we obtained. In the following, we use some snippets extracted from ERC20 examples in our evaluation to illustrate the application of our framework, some of the errors that we found, and a case of a safe evolution.

The ERC20 is likely to be the most widely implemented Ethereum standard. It defines member variables: \lstinline|totalSupply| keeps track of the total number of tokens in circulation, \lstinline|balanceOf| maps a wallet (i.e. address) to the balance it owns, and \lstinline|allowance| stores the number of tokens that an address has made available to be spent by another one. It defines public functions: \lstinline|totalSupply|, \lstinline|balanceOf| and \lstinline|allowance| are accessors for the above variables; \lstinline|transfer| and \lstinline|transferFrom| can be used to transfer tokens between contracts; and \lstinline|approve| allows a contract to set an ``allowance" for a given address.

\begin{figure}[!t]
\begin{lstlisting}[basicstyle=\scriptsize\ttfamily, numbers=left, escapechar=`]
/// @notice invariant totalSupply == __verifier_sum_uint(balanceOf) `\label{line:invariant}`
contract IERC20  {
`\label{line:datastructurebegin}`uint256 totalSupply;
mapping (address => uint256) balanceOf;
mapping (address => mapping (address => uint256)) allowance;`\label{line:datastructureend}`

//... functions transfer, totalSupply, balanceOf, and approve omitted ...

`\label{line:postconditionallowance}`/// @notice postcondition allowance[owner][spender] == remaining
function allowance(address owner, address spender) external view returns (uint256 remaining);

/**
`\label{line:postconditionbalancebegin}`* @notice postcondition ( ( balanceOf[_from] == __verifier_old_uint ( balanceOf[_from] ) - _value && _from != _to )  ||  ( balanceOf[_from] == __verifier_old_uint ( balanceOf[_from] ) && _from == _to ) && success )  || !success
* @notice postcondition ( ( balanceOf[_to] == __verifier_old_uint ( balanceOf[_to] ) + _value && _from != _to ) ||  ( balanceOf[_to] == __verifier_old_uint ( balanceOf[_to] ) && _from == _to ) && success )  || !success `\label{line:postconditionbalanceend}`   
`\label{line:postconditionallowancebegin}`* @notice postcondition allowance[_from][msg.sender] == __verifier_old_uint ( allowance[_from][msg.sender] ) - _value || _from == msg.sender 
* @notice postcondition allowance[_from][msg.sender] <= __verifier_old_uint ( allowance[_from][msg.sender] ) || _from == msg.sender `\label{line:postconditionallowanceend}`
*/
function transferFrom(address _from, address _to, uint256 _value) public returns (bool success);`\label{line:transferfromend}`
}	
\end{lstlisting}
\caption{ERC20 reduced specification}
\label{fig:erc20_spec}
\end{figure}

Figure~\ref{fig:erc20_spec} presents a reduced specification, focusing on functions \lstinline|transferFrom| and \lstinline|allowance| for the purpose of this discussion, derived from the informal description in the standard~\cite{erc20}. In Line~\ref{line:invariant}, we define an invariant requiring that the total number of tokens remain unchanged regardless of the operation carried out by the contract. The function \lstinline|allowance|
does not change the state of the smart contract so it has only one
postcondition (line~\ref{line:postconditionallowance}) to ensure that it will return the correct amount of tokens available for withdrawals. The \lstinline|transferFrom| function has 4 postconditions;  the  operation  is  successful  only  when  the  tokens  are  debited  from the source account and credited in the destination account, according to the specifications  provided  in  the  ERC20  standard. The first two postconditions (lines \ref{line:postconditionbalancebegin} to \ref{line:postconditionbalanceend}) require that the balances are updated as expected, whereas the purpose of the last two (lines \ref{line:postconditionallowancebegin} to \ref{line:postconditionallowanceend}) is to ensure  that the tokens available for withdrawal have been properly updated.

We use the snippet in Figure~\ref{fig:buggytransfermethod} - extracted from the Uniswap repository, commit \underline{55ae25}, we underline the commits that we refer to in Table 1 as well to help readers locate it more easily - to illustrate the detection of wrong operator errors. When checked by our framework, the third postcondition for the \lstinline|transferFrom| function presented in the specification in Figure~\ref{fig:erc20_spec} is not satisfied. Note that the allowance amount is not debited if the amount to be transferred is equal to the maximum integer supported by Solidity (i.e. \lstinline|uint(-1)|). A possible solution would consist of removing the \lstinline|if| branching, allowing the branch code to always execute.
  
\begin{figure}[!t]
\begin{lstlisting}[basicstyle=\scriptsize\ttfamily, escapechar=`]
function transferFrom(address from, address to, uint value) external 
returns (bool success) { 
    if (allowance_[from][msg.sender] != uint(-1)) { 
        allowance_[from][msg.sender] = 
        allowance_[from][msg.sender].sub(value); 
    } 
    _transfer(from, to, value); 
    return true; 
}
\end{lstlisting}
\caption{Buggy ERC20 transferFrom function}
\label{fig:buggytransfermethod}
\end{figure}

The code snippet in Figure \ref{fig:allowancemethod} - DigixDao repository, commit \underline{5aee64} - does not conform to its formal specification. The correct allowance for the spender is only returned when it is not greater than the owner's balance. To fix this issue, we need to remove all code related to \lstinline|_balance|, ensuring that the \lstinline|_allowance| will be returned regardless of the \lstinline|_balance| amount.

\begin{figure}[!b]
\begin{lstlisting}[basicstyle=\scriptsize\ttfamily, escapechar=`]
 function allowance(address _owner, address _spender) public returns (uint256 remaining) {
     uint256 _allowance = allowed[_owner][_spender];
     uint256 _balance = balances[_owner];
     if (_allowance > _balance) {
        remaining = _balance;
     } else {
        remaining =  _allowance;
     }
     return remaining;
} 
\end{lstlisting}
\caption{Buggy ERC20 allowance function}
\label{fig:allowancemethod}
\end{figure}

The ERC1155 (see Figure~\ref{fig:erc1155specification}) was created in order to promote a better integration between the ERC20 and ERC721 standards. It provides an interface for managing any combination of fungible and non-fungible tokens in a single contract efficiently. The functions \lstinline|balanceOf| and \lstinline|balanceOfBatch| returns the balance of specific tokens of the address or a list of addresses specified in the parameter function respectively. The function \lstinline|isApprovedForAll| returns a boolean value informing if an address is allowed to handle the tokens from another address. The \lstinline|safeTransferFrom| function transfers tokens in a safe way to a valid ERC1155 address. The \lstinline|transferFrom| function can do batch operations, transfering tokens to several wallets at the same time, reducing transaction costs and minimizing impacts on the network. The \lstinline|setApprovalForAll| function gives an address permission to handle another address' tokens.

\begin{figure}[!t]
\begin{lstlisting}[basicstyle=\scriptsize\ttfamily, escapechar=`]
contract ERC1155 {
    
    mapping (uint256 => mapping(address => uint256)) private _balances;
    mapping (address => mapping(address => bool)) private _operatorApprovals;

    /** @notice postcondition _balances[id][account] == balance */
    function balanceOf(address account, uint256 id) public view   returns (uint256 balance) {
    }
     /** @notice postcondition batchBalances.length == accounts.length 
    * @notice postcondition batchBalances.length == ids.length
    * @notice postcondition forall (uint x) !( 0 <= x &&  x < batchBalances.length ) || batchBalances[x] == _balances[ids[x]][accounts[x]] */
    function balanceOfBatch( address[] memory accounts, uint256[] memory ids ) public view returns (uint256[] memory batchBalances) {
    }
    /** @notice  postcondition _operatorApprovals[msg.sender][operator] ==  approved 
    * @notice  emits  ApprovalForAll */
    function setApprovalForAll(address operator, bool approved) public {
    }
    /** @notice postcondition _operatorApprovals[account][operator] == approved */
    function isApprovedForAll(address account, address operator) public view   returns (bool approved) {
        return _operatorApprovals[account][operator];
    }
    /** @notice postcondition to != address(0)
    * @notice postcondition _operatorApprovals[from][msg.sender] || from == msg.sender
    * @notice postcondition __verifier_old_uint ( _balances[id][from] ) >= amount    
    * @notice postcondition _balances[id][from] == __verifier_old_uint ( _balances[id][from] ) - amount
    * @notice postcondition _balances[id][to] == __verifier_old_uint ( _balances[id][to] ) + amount
    * @notice emits TransferSingle  */
    function safeTransferFrom(address from, address to, uint256 id, uint256 amount, bytes memory data ) public {
    }
    /** @notice postcondition _operatorApprovals[from][msg.sender] || from == msg.sender
    * @notice postcondition to != address(0)
    * @notice emits TransferBatch */
    function safeBatchTransferFrom( address from, address to,  uint256[] memory ids, uint256[] memory amounts, bytes memory data) public {
    }
}
\end{lstlisting}
\caption{ERC1155 specification}
\label{fig:erc1155specification}
\end{figure}

The snippet in Figure~\ref{fig:buggysafeBatchTransferFrom} - extracted from Descentralized-Stock repository, commit \underline{96d5b2} - illustrates another example of the wrong operator error. A postcondition was not satisfied, because, according to the specification, the size of the \lstinline|_ids| and \lstinline|_values| arrays must be equal. So, any call to this function would result in an error or an unexpected behavior. A possible solution would consist of changing the operator \lstinline|!=| for \lstinline|==| in the second \lstinline|requires| in Figure~\ref{fig:buggysafeBatchTransferFrom}.

Figures \ref{fig:refactoringmethodbefore} and \ref{fig:refactoringtransfermethod} - extracted from the 0xSequence-erc-1155 repository, 
commits \underline{99012f} and \underline{319740}, respectively - illustrate a case of safe contract evolution. The code of this contract has undergone  significant changes. The refactoring in question is one of the most common and is known as extract method (function, in Solidity). From commit \underline{319740} to \underline{99012f}, the new internal function \lstinline|_callonERC1155Received| was created, and the extracted code from the \lstinline|_safeTransferFrom| was moved into it. 

\begin{figure}[!t]
\begin{lstlisting}[basicstyle=\scriptsize\ttfamily, escapechar=`]
function safeTransferFrom(address _from, address _to, uint256 _id, uint256 _value, bytes memory _data) public {
    
    require((msg.sender == _from) || operators[_from][msg.sender], "INVALID_OPERATOR");
    require(_to != address(0),"INVALID_RECIPIENT");
    require(_value >= balances[_from][_id]) is not necessary since checked with safemath operations

    _safeTransferFrom(_from, _to, _id, _value, _data);
  }
\end{lstlisting}
\caption{safeTransferFrom function before refactoring}
\label{fig:refactoringmethodbefore}
\end{figure}

\begin{figure}[!b]
\begin{lstlisting}[basicstyle=\scriptsize\ttfamily, escapechar=`]
function safeTransferFrom(address _from, address _to, uint256 _id, uint256 _amount, bytes memory _data) public {
    
    require((msg.sender == _from) || operators[_from][msg.sender], "ERC1155#safeTransferFrom: INVALID_OPERATOR");
    require(_to != address(0),"ERC1155#safeTransferFrom: INVALID_RECIPIENT");
    require(_amount >= balances[_from][_id]) is not necessary since checked with safemath operations

    _safeTransferFrom(_from, _to, _id, _amount);
    _callonERC1155Received(_from, _to, _id, _amount, _data);
  }
\end{lstlisting}
\caption{Successful refactoring of the safeTransferFrom function}
\label{fig:refactoringtransfermethod}
\end{figure}

\begin{figure}[!b]
\begin{lstlisting}[basicstyle=\scriptsize\ttfamily, escapechar=`]
  function safeBatchTransferFrom(address _from, address _to, uint256[] calldata _ids, uint256[] calldata _values, bytes calldata _data) external {
        require(_to != address(0) && _from != address(0));
        require(_ids.length != _values.length);
        require(_approv[_from][msg.sender] || _from == msg.sender);

        for (uint256 i = 0; i < _ids.length; ++i) {
            require(_balance[_from][_ids[i]] >= _values[i]);
            _balance[_from][_ids[i]] -= _values[i];
            _balance[_to][_ids[i]] += _values[i];
        }
        emit TransferBatch(msg.sender, _from, _to, _ids, _values);
        require(_checkOnERC1155BatchReceived(msg.sender, _from, _to, _ids, _values, _data));
    }
\end{lstlisting}
\caption{Buggy ERC1155 safeBatchTransferFrom function}
\label{fig:buggysafeBatchTransferFrom}
\end{figure}

The results of our evaluation suggest that the kind of verification that we employ in our framework is tractable, as \emph{solc-verify} can efficiently analyse these samples. The fact that errors that could lead to millionaire losses were detected in real-world contracts attests to the necessity of our framework and its practical impact. Smart contracts are incresingly popular, and we believe they will become a key and common element of trusted distributed systems in the future. Therefore, having a safe development process supported by our framework will help to increase the credibility of such a technology and promote its adoption.

\subsection{Threats to Validity and Limitations}
\label{sec:threatslimitations}

Our initial motivation to gather the samples from public github repositories may be a threat to our search strategy. Since we could not analyze private repositories, relevant cases to show strengths and weaknesses of the framework may not have been included. The relatively low number of samples could also be considered as a threat, since it could lead to an unintentionally biased study or less comprehensive than it could have been. Our framework also presents some limitations, since it is not in our scope to verify errors on inter-contract interactions.

\subsection{Trusted Deployer Tool}
\label{sec:upgradertool}

After discussing the results of the smart contract verification process, we introduce a scenario based on 0xMonorepo repository which implements the ERC20 pattern and is one of the repositories used during our study. We analysed the whole commit history (see Table~\ref{tab:0xmonorepo}) in order to find errors or nonconformities and describe what the history of the repository would be if a developer had used our tool since invalid commits would be prevented from being deployed. The idea is to show how our tool implements the architecture depicted in Figure~\ref{fig:trusted-deployer}; we also demonstrate the tool supports a safe smart contract development process.

As already explained, the trusted deployer requires that developers have their code formally verified before they can deploy their contracts to a blockchain network. In order to create or upgrade a smart contract, a developer has to provide its code and specification together. The tool uses the solc-verify in background to verify the code against its specification before it proceeds to deploy the smart contract. Figure ~\ref{fig:reducedmergedcontract} presents a merged contract, which is the result of the merging of the specification and implementation contracts. The verification contract is automatically created from the abstract syntax tree of the contracts after a syntactic check is performed; in this case, the goal is to analyse if there is any discrepancy between the signature of the functions and the data model of the specification and implementation contracts. Provided the syntactic analysis is successful, the tool invokes the solc-verify in background to carry out verification of conformance to the specification.

\begin{figure}[H]
\begin{lstlisting}[basicstyle=\scriptsize\ttfamily,  escapechar=`]
/// @notice  invariant  _totalSupply  ==  __verifier_sum_uint(balances)
contract ERC20Token {

    uint constant MAX_UINT = 2**256 - 1;
    mapping (address => uint) balances;
    mapping (address => mapping (address => uint)) allowed;
    uint public _totalSupply;
    event Transfer(address indexed _from, address indexed _to, uint _value);
    event Approval(address indexed _owner, address indexed _spender, uint _value);
    
   
\end{lstlisting}
\end{figure}

\begin{figure}[H]
\begin{lstlisting}[basicstyle=\scriptsize\ttfamily,  escapechar=`]
    /** @notice  postcondition (( balances[msg.sender] ==  __verifier_old_uint (balances[msg.sender]) - _value  && msg.sender  != _to) || (balances[msg.sender] ==  __verifier_old_uint ( balances[msg.sender]) && msg.sender  == _to) &&  success) || !success
    * @notice  postcondition (( balances[_to] ==  __verifier_old_uint (balances[_to]) + _value  && msg.sender  != _to) || (balances[_to] ==  __verifier_old_uint (balances[_to]) && msg.sender  == _to )) || !success
    * @notice  emits  Transfer */
    function transfer(address _to, uint _value) public returns (bool success) {
        require(balances[msg.sender] >= _value && balances[_to] + _value >= balances[_to]); 
        balances[msg.sender] -= _value;
        balances[_to] += _value;
        emit Transfer(msg.sender, _to, _value);
        return true;
    }

    /** @notice  postcondition (( balances[_from] ==  __verifier_old_uint (balances[_from] ) - _value  &&  _from  != _to) || ( balances[_from] ==  __verifier_old_uint ( balances[_from] ) &&  _from == _to ) && success ) || !success 
    * @notice  postcondition ( ( balances[_to] ==  __verifier_old_uint ( balances[_to] ) + _value  &&  _from  != _to ) || ( balances[_to] ==  __verifier_old_uint ( balances[_to] ) &&  _from  == _to ) && success ) || !success 
    * @notice  postcondition ( allowed[_from ][msg.sender] ==  __verifier_old_uint (allowed[_from ][msg.sender] ) - _value && success) || ( allowed[_from ][msg.sender] ==  __verifier_old_uint (allowed[_from ][msg.sender]) && !success) ||  _from  == msg.sender
    * @notice  postcondition  allowed[_from ][msg.sender]  <= __verifier_old_uint (allowed[_from ][msg.sender] ) ||  _from  == msg.sender
    * @notice  emits  Transfer */
    function transferFrom(address _from, address _to, uint _value) public  returns (bool success) {
        uint allowance = allowed[_from][msg.sender];
        require(balances[_from] >= _value && allowance >= _value && balances[_to] + _value >= balances[_to]); 
        balances[_to] += _value;
        balances[_from] -= _value;
        if (allowance < MAX_UINT) {
            allowed[_from][msg.sender] -= _value;
        }
        emit Transfer(_from, _to, _value);
        return true;
    }

    /** @notice  postcondition (allowed[msg.sender ][ _spender] ==  _value  &&  success) || ( allowed[msg.sender ][ _spender] ==  __verifier_old_uint ( allowed[msg.sender ][ _spender] ) && !success )    
    * @notice  emits  Approval */
    function approve(address _spender, uint _value)  public returns (bool success) {
        allowed[msg.sender][_spender] = _value;
        emit Approval(msg.sender, _spender, _value);
        return true;
    }

    /** @notice postcondition balances[_owner] == balance */
    function balanceOf(address _owner) public view returns (uint balance) {
        return balances[_owner];
    }

    /** @notice postcondition allowed[_owner][_spender] == remaining */
    function allowance(address _owner, address _spender) public view returns (uint remaining) {
        return allowed[_owner][_spender];
    }
}
\end{lstlisting}
\caption{Merged Contract}
\label{fig:reducedmergedcontract}
\end{figure}

The trusted registry (see Figure~\ref{fig:truestedregistry}) check whether that a given instance was created by the trusted deployer by calling its \textit{get-spec} function. Since smart contracts in a blockchain platform cannot rely on external services, such as the trusted deployer, that is, they would have no means to check whether an instance that they want to interact with is safe or not. We create a trusted registry as part of the trusted deployer infrastructure which is essentially a mirror of the trusted deployer’s internal registry implemented as a smart contract. It has a maintainer and mapping \lstinline|verified_addr| that associates the proxy instances that have been created by the trusted deployer with the specification they comply to. As an implementation detail, we do not store the specification themselves but rather a small representative as a 32-byte array - it could be, for instance, a cryptographic hash of the specification. We do not allow this representative to be the zeroed array, $bytes32(0)$ in Solidity syntax, as we use this value to represent absence of an association. That is, if the result of a call $get\_spec(addr)$ is the value $bytes32(0)$, it means the address $addr$ has not been deployed by the trusted deployed, and hence has no specification associated with it.

\begin{figure}[H]
\begin{lstlisting}[basicstyle=\scriptsize\ttfamily, numbers=left, escapechar=`]
contract Registry {
    address maintainer;
    mapping (address => bytes32) verified_addrs;

    constructor() public {
        maintainer = msg.sender;
    }
    function new_mapping(address addr, bytes32 spec_id) public {
        if (msg.sender == maintainer && spec_id != bytes32(0)) {
            verified_addrs[addr] = spec_id;
        }
    }
    function get_spec(address addr) view public returns (bytes32) {
        return verified_addrs[addr];
    }
}    
\end{lstlisting}
\caption{Trusted registry}
\label{fig:truestedregistry}
\end{figure}

The proxy pattern as discussed in Section~\ref{sec:upgrader} separates the logic and the data of a contract. The key concept to understand is that the implementation contract can be replaced while the  trusted proxy (see Figure~\ref{fig:reducedImplementedproxy}), or the access point is never changed. 
Both contracts are still immutable in the sense that their code cannot be changed, but the logic contract can simply be swapped by another contract. Every proxy must contain all variables (lines \ref{line:variablesproxybegin} to \ref{line:variablesproxyend}) defined in the implementation contract which will store the state as well as the signatures of all its public functions (lines \ref{line:functionproxybegin} to \ref{line:functionproxyend}). If there is any constructor in the implementation contract it will be merged with the default constructor of the proxy (lines \ref{line:constructorproxybegin} to \ref{line:constructorproxyend}).

\begin{table}[!t]
\setlength{\tabcolsep}{3pt}
\resizebox{\textwidth}{!}{%
\begin{tabular}{|c|c|c|c||c|c|c|c|}
\hline
 \multicolumn{8}{|c|}{0xMonorepo Repository} \\
\hline
 Commit & Date & Commit & Date & Commit & Date & Commit & Date\\
\hline
	\underline{7d59fa} & 12/12/2017 &	bb4c8b & 02/02/2018 &	89abd7 & 18/05/2018 & 99fbf3 & 04/08/2018\\				
	\underline{7008e8} & 12/12/2017 &	897515 & 06/02/2018 &	ba1485 & 22/06/2018 & 9b521a & 20/12/2018\\			
	\underline{b58bf8} & 12/12/2017 &	32fead & 21/03/2018 &	f21b04 & 05/07/2018 & 0758f2 & 22/01/2019\\				
	\underline{548fda} & 12/12/2017 &	d11811 & 21/03/2018 &	d2e422 & 06/07/2018 & d35a05 & 07/03/2019\\				
	6f2cb6 & 12/12/2017 &	1729cf & 20/04/2018 &	a2024d & 06/07/2018 & 5813bb & 23/07/2019\\				
	145fea & 12/12/2017 &	63abf3 & 21/04/2018 &	bb3c34 & 16/08/2018 & 01aeee & 21/10/2019\\				
	1fb643 & 12/12/2017 &	c84be8 & 02/05/2018 &	f54591 & 16/08/2018 & & \\				
	272125 & 12/12/2017 &	5198c5 & 08/05/2018 &	8bce73 & 16/08/2018 & & \\
    
 \hline
\end{tabular}}
\caption{0xMonorepo Commit History}
\label{tab:0xmonorepo}
\end{table}

During the analysis of the scenario 30 commits were verified, it was observed the WOP error in its first four commits \underline{7d59fa}, \underline{7008e8}, \underline{b58bf8} and \underline{548fda}. If the developer had used our tool the error would have been discovered in the first analysis, these deployments would have been prevented, and an error message containing the specific reason would be returned to the developer forcing him to fix the bug. The results collected from our evolution scenario, one can see that our strategy is effective in identifying error in the early stage of the process. Our tool abstracts many details of the deployment and upgrade process making it simpler for platform users when compared to the manual process.

\begin{figure}[H]
\begin{lstlisting}[basicstyle=\scriptsize\ttfamily,    escapechar=`]
contract Proxy {

 `\label{line:variablesproxybegin}`   uint256 public _totalSupply; 
    mapping(address => uint256) internal balances; 
    mapping(address => mapping(address => uint256)) internal allowed; `\label{line:variablesproxyend}` 

  `\label{line:functionproxybegin}`  function approve (address _spender,uint256 _value) public returns (bool success) {
        (bool success, bytes memory bytesAnswer) = implementation.delegatecall(abi.encodeWithSignature("approve(address,uint256)",_spender,_value));
            require(success);
            return abi.decode(bytesAnswer, (bool));
    }
    
    function transfer (address _to,uint256 _value) public returns (bool success) {
        (bool success, bytes memory bytesAnswer) = implementation.delegatecall(abi.encodeWithSignature("transfer(address,uint256)" ,_to,_value));
            require(success);
            return abi.decode(bytesAnswer, (bool));
    }
    
    function transferFrom (address _from,address _to,uint256 _value) public returns (bool success) {
        (bool success, bytes memory bytesAnswer) = implementation.delegatecall(abi.encodeWithSignature("transferFrom(address,address,uint256)" ,_from,_to,_value));
                    require(success);
                     return abi.decode(bytesAnswer, ( bool ) );
    }
\end{lstlisting}
\end{figure}

\begin{figure}[H]
\begin{lstlisting}[basicstyle=\scriptsize\ttfamily,    escapechar=`]
     function balanceOf (address _owner) public returns (uint256 balance) {
    (bool success, bytes memory bytesAnswer) = implementation.delegatecall(abi.encodeWithSignature("balanceOf(address)" ,_owner));
                    require(success);
                     return abi.decode(bytesAnswer, ( uint256 ) );
    }
    function allowance (address _owner,address _spender) public   returns (uint256 remaining) {
        (bool success, bytes memory bytesAnswer) = implementation.delegatecall(abi.encodeWithSignature("allowance(address,address)" ,_owner,_spender));
            require(success);
            return abi.decode(bytesAnswer, ( uint256 ) );
    }`\label{line:functionproxyend}`

    Registry registry;
    bytes32 spec;
    address implementation;
    address author;

    `\label{line:constructorproxybegin}`constructor(Registry _registry, bytes32 _spec, address _implementation ) public {
        require(_spec != bytes32(0));
        registry = _registry;
        spec = _spec;
        author = msg.sender;
        _upgrade(_implementation);
    }`\label{line:constructorproxyend}`
    function upgrade(address new_implementation) public {
        _upgrade(new_implementation);
    }
    function _upgrade(address new_implementation) internal {
        require(msg.sender == author);
        bytes32 spec_id = registry.get_spec(new_implementation);
        require(spec_id == spec);
        implementation = new_implementation;
    }
}
\end{lstlisting}
\caption{Trusted Proxy}
\label{fig:reducedImplementedproxy}
\end{figure}

\section{Related Work}
\label{sec:related-work}

There is a glaring need for a safe mechanism to upgrade smart contracts in platforms, such as Ethereum, where contract implementations are immutable once deployed; the many surveys uncovering this fact~\cite{Hu21,Tolmach22,Groce20} and community-proposed design patterns proposing mechanisms to upgrade smart contracts \cite{proxy,uups,delegateproxy,spss} attest this necessity. Yet, surprisingly, we could only find two close related approaches~\cite{Dickerson18,Rodler21} that try to tackle this problem. The preliminary work in~\cite{Dickerson18} proposes a methodology based around special contracts that carry a proof that they meet the expected specification. Their on-chain solution requires fundamental changes in the smart contract platforms themselves. They propose the addition of a special instruction to deploy these special proof-carrying contracts, and the adaptation of platform miners, which are responsible for checking and reaching a consensus on the validity of contract executions, to check these proofs. Our framework and the one presented in that work share the same goal: to propose a mechanism by which contracts can be upgraded but only if they meet the expected specification. However, our approach and theirs differ significantly in many aspects. Firstly, while theirs requires a fundamental change on the rules of the platform - which requires a large distributed network of nodes, i.e. the smart contract platform, to agree upon - ours can be implemented, as already prototyped, on top of Ethereum's current capabilities and can rely on tools that are easier to use, i.e. require less user input, like program verifiers. The fact that their framework is on-chain makes the use of such verification methods more difficult since these methods would slow down consensus, likely to a prohibitive level. Finally, while they introduce abstract ideas with some concrete elements, we provide details on how to implement our framework using current technology and an evaluation based on real-world Solidity samples.

In~\cite{Rodler21}, the authors propose a mechanism to upgrade contracts in Ethereum that works at the EVM-bytecode level. Their framework takes vulnerability reports issued by the community as an input, and tries to patch affected deployed contracts automatically using patch templates. It uses previous contract transactions and, optionally user-provided unit tests, to try to establish whether a patch preserves the behaviour of the contract. Ultimately, the patching process may require some manual input. If the deployed contract and the patch disagree on some test, the user must examine this discrepancy and rule on what should be done. Note that this manual intervention is always needed for attacked contracts, as the transaction carrying out the attack - part of the attacked contract's history - should be prevented from happening in the new patched contract.
While they simply test patches that are reactively generated based on vulnerability reports, we proactively require the user to provide a specification of the expected behaviour of a contract and formally verify the evolved contract against such a formal specification. Their approach requires less human intervention, as a specification does not need to be provided - only optionally some unit tests - but it offers no formal guarantees about patches. It could be that a patch passes their validation (i.e. testing with the contract history), without addressing the underlying vulnerability. 

Some other papers have proposed methodologies to carry out pre-deployment patching/repairing~\cite{Torres21,Nguyen21,Yu20}. They try to scan a binary for common vulnerabilities and patch the vulnerabilities they find prior to deploying the contract. These papers do not propose a way to update deployed contracts.

A number of analysis tools for EVM bytecode were designed to find specific behaviour patterns witnessing typical bad behaviours~\cite{Luu16,Mossberg19,Liu18,Grishchenko18a,Tikhomirov18,Tsankov18,Permenev20}. Tools operating on the level of Solidity were also proposed~\cite{Wang18,Hajdu19,Hajdu20,Antonino21,Ahrendt19,Ahrendt20}. These tools tend to focus instead on formally verifying user-provided semantic properties. Our paper proposes a verification-focused development process based around, supported, and enforced by such tools.

Design by contract~\cite{Meyer92} is a methodology that was originally created for specifying the behaviour of object-oriented programs but was also adopted in other contexts~\cite{Meyer88,Leino10,Barnett05,Leino08,Leavens99,Hajdu19}. This sort of specification is particularly fitting in the case of Solidity smart contracts, especially the format of specification that we propose, as the community already employ a similar format, albeit informal, to describe standard contract interfaces in the form of Ethereum Request for Comments (ERCs); see for example, ERC20~\cite{erc20}.

\section{Conclusion}
\label{sec:conclusion}

We propose a framework for the 
safe deployment of smart contracts.
Not only does it check that contracts conform to their specification at creation time, but it also guarantees that subsequent code updates are conforming too. Upgrades can be performed even if the implementation has been proven to satisfy the specification initially. A developer might, for instance, want to optimise the resources used by the contract. Furthermore, our \emph{trusted deployer} 
records information about the contracts that have been verified, and which specification they conform to, so that participants can be certain they are interacting with a contract with the expected behaviour; contracts can be safely executed. None of these capabilities are offered by the Ethereum platform by default nor are available in the literature to the extent provided by the framework proposed in this paper.

We have prototyped our trusted deployer and investigated its applicability - specially its formal verification component - to contracts implementing two widely used Ethereum standards: the ERC20 Token Standard and ERC1155 Multi Token Standard, with promising results. 

This idea of using trusted computing to verify a smart contract before its deployment can be extended to software in general. Particularly, a trusted deployer could be part of a deployment process for reactive systems in general, such as component-based, (micro)service-based systems, or even system-of-systems.

Our framework shifts immutability from the implementation of a contract to its specification, promoting the ``code is law" to the ``specification is law" paradigm. We believe that this paradigm shift brings a series of improvements. Firstly, developers are required to outline their intent in the form of a (formal) specification, so they can, early in the development process, identify issues with their design. They can and should validate their specification; we consider this problem orthogonal to the framework that we are providing. Secondly, specifications are more abstract and, as a consequence, tend to be more stable than (the corresponding conforming) implementations. A contract can be optimised, for instance, and both the original and optimised versions must satisfy the same reference specification. Thirdly, even new implementations that involve change of data representation can still be formally verified against the same specification, by using data refinement techniques. 

A limitation of our current approach is the restrictive notion of evolution for smart contracts: only the implementation of public functions can be upgraded - the persistent state data structures are fixed. However, we are looking into new types of evolution where the data structure of the contract's persistent state can be changed - as well as the interface of the specification, provided the projected behaviour with respect to the original interface is preserved, based on notions of class~\cite{Liskov94} and process~\cite{Dihego13} inheritance, and interface evolution such as in~\cite{Dihego20}.

\bibliographystyle{plain}
\bibliography{references}
\end{document}